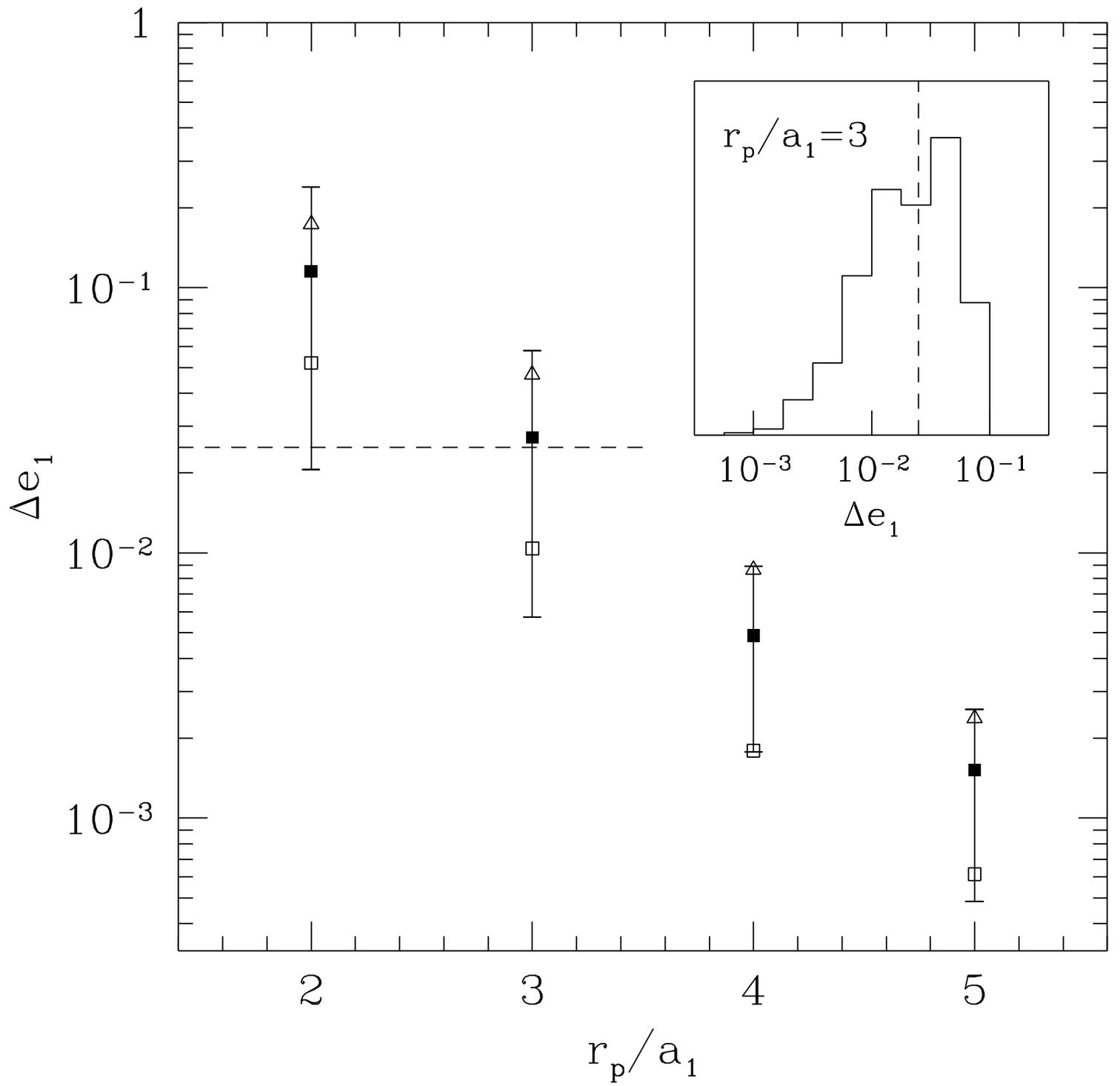

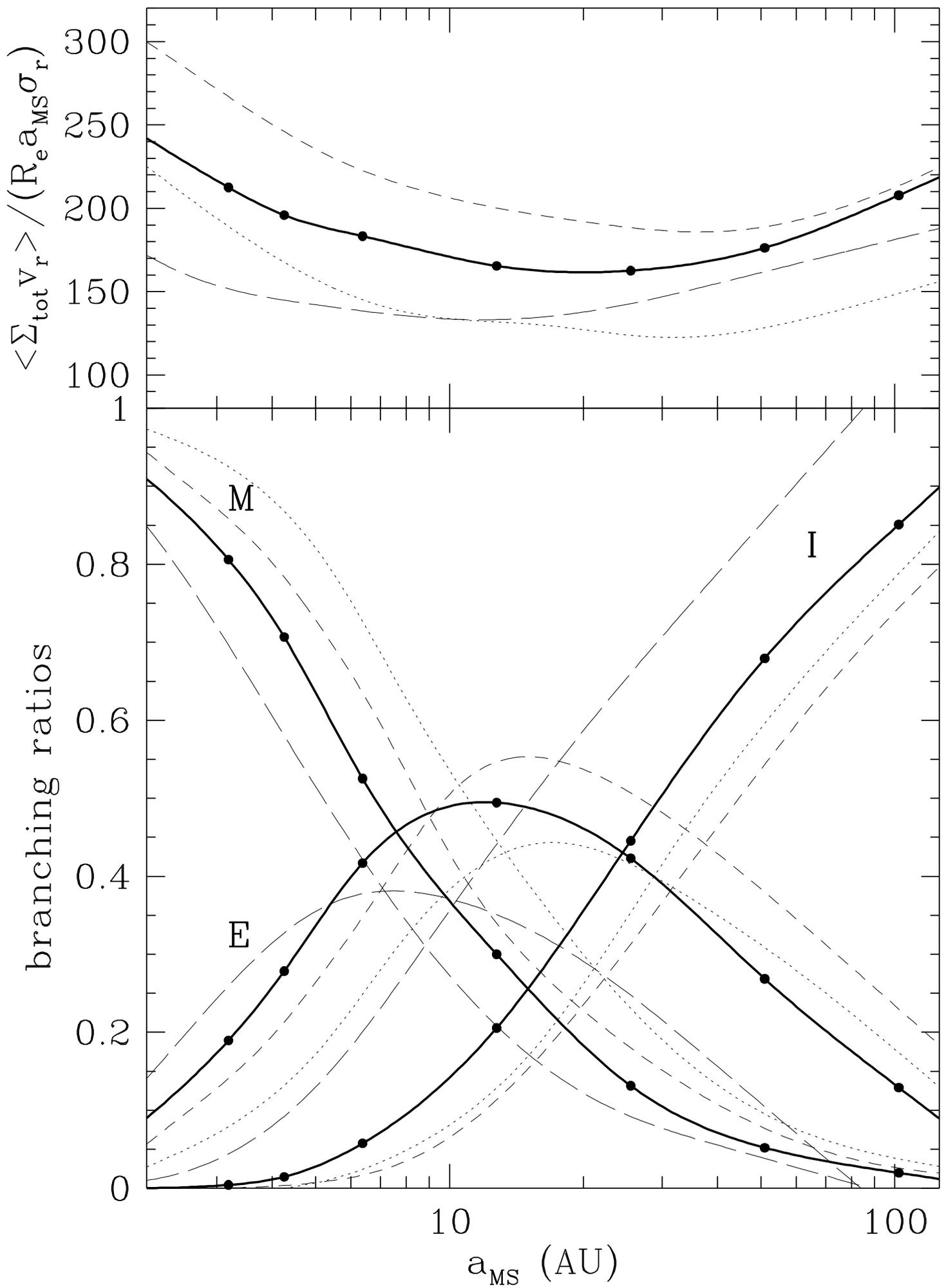

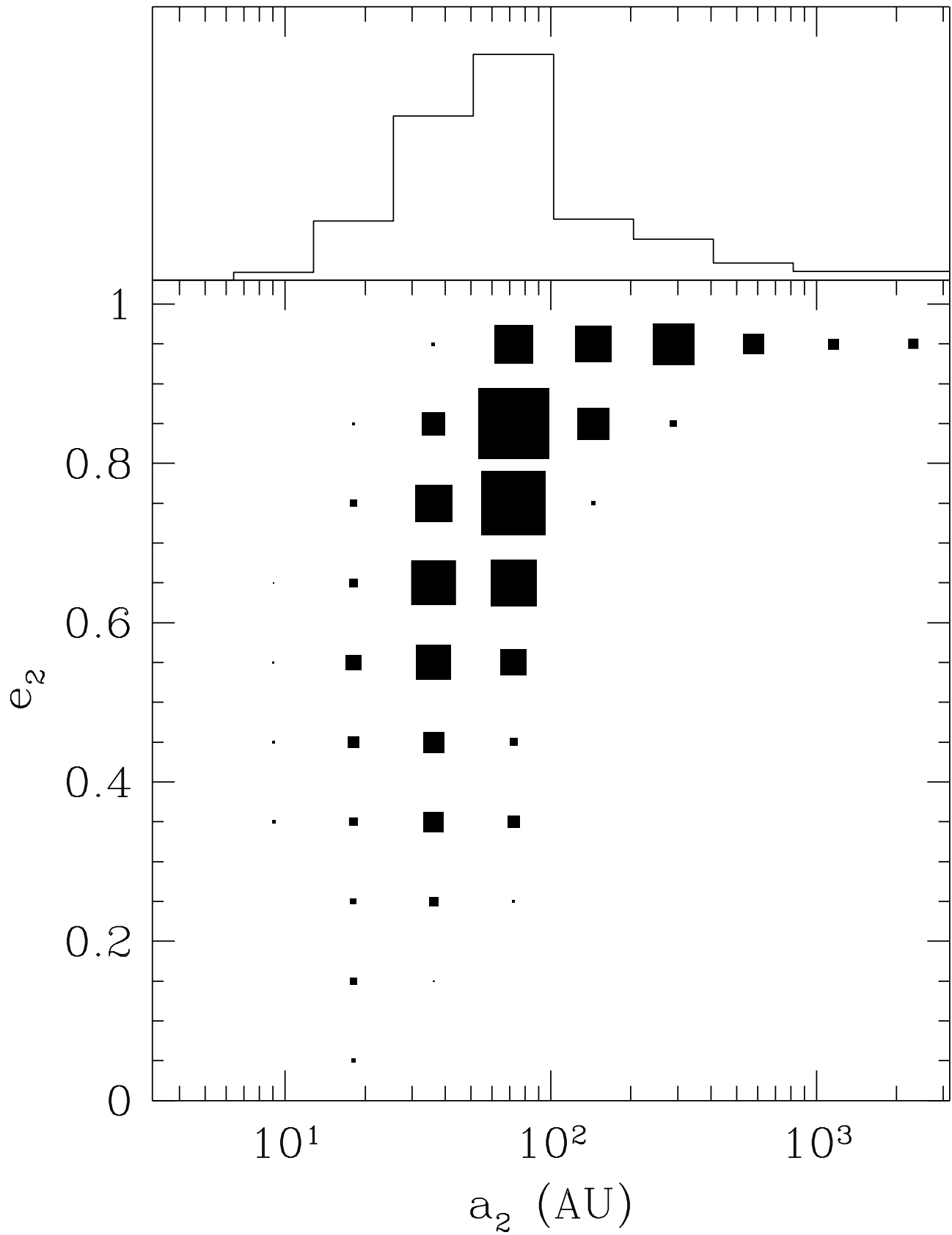

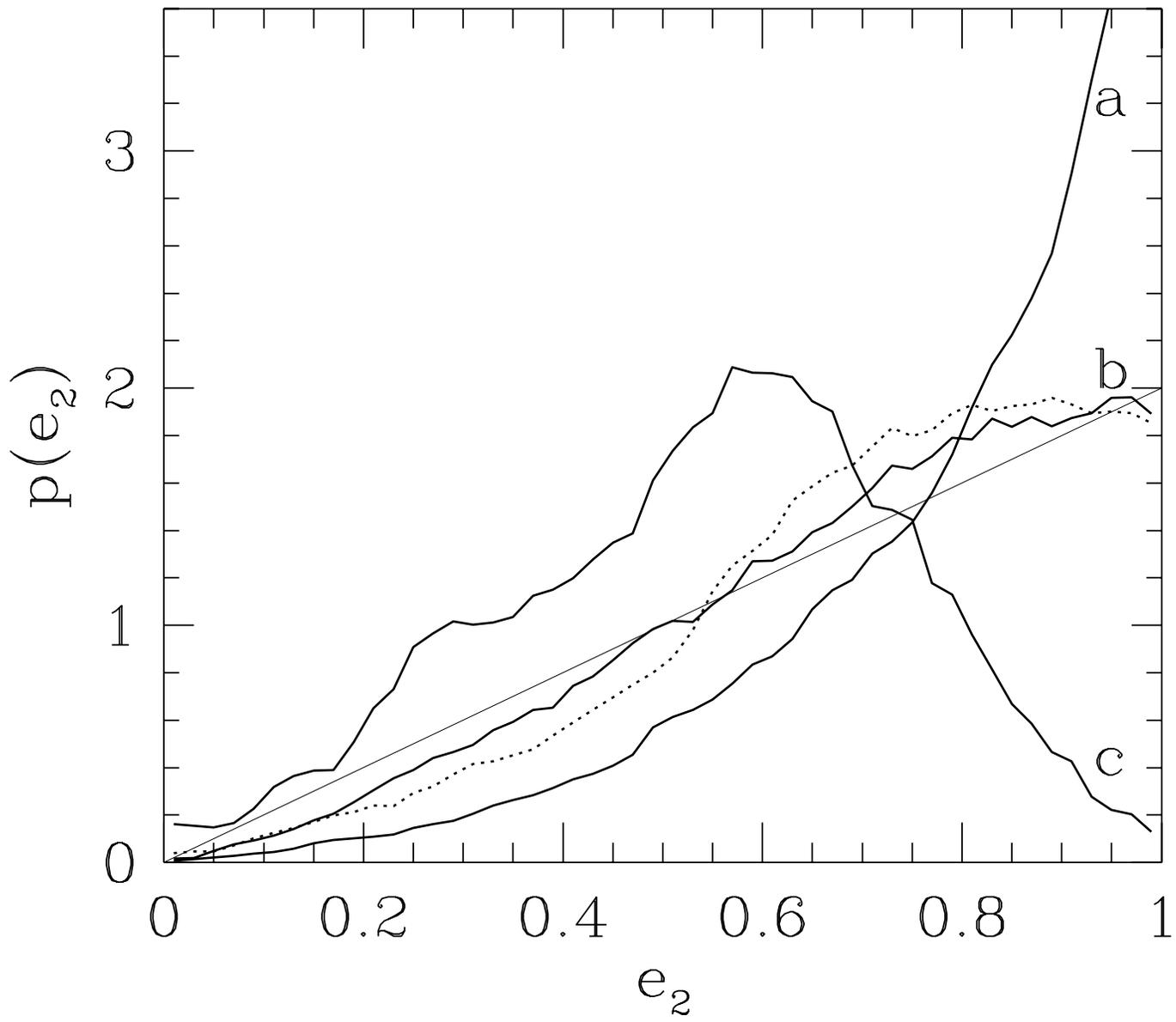

# BINARY-BINARY INTERACTIONS AND THE FORMATION OF THE PSR B1620−26 TRIPLE SYSTEM IN M4

Frederic A. Rasio[1,2], Steve McMillan[3], and Piet Hut[1]

## ABSTRACT

The hierarchical triple system containing the millisecond pulsar PSR B1620−26 in M4 is the first triple star system ever detected in a globular cluster. Such systems should form in globular clusters as a result of dynamical interactions between binaries. We propose that the triple system containing PSR B1620−26 formed through an exchange interaction between a wide primordial binary and a *pre-existing* binary millisecond pulsar. This scenario would have the advantage of reconciling the $\sim 10^9$ yr timing age of the pulsar with the much shorter lifetime of the triple system in the core of M4.

*Subject headings:* celestial mechanics, stellar dynamics — globular clusters: general — globular clusters: individual (M4) — pulsars: individual (PSR B1620-26)

---

[1]Institute for Advanced Study, Princeton, NJ 08540.

[2]Hubble Fellow.

[3]Department of Physics and Atmospheric Science, Drexel University, Philadelphia, PA 19104.

## 1. Introduction

The millisecond pulsar PSR B1620−26 in the globular cluster M4 has a low-mass companion (most likely a white dwarf, of mass $m_1 \approx 0.3\,M_\odot$ for a pulsar mass $m_p = 1.35\,M_\odot$) in a nearly circular orbit of period $P_1 = 0.524\,\mathrm{yr}$ (Lyne et al. 1988; McKenna & Lyne 1988), corresponding to a separation $a_1 \approx 0.8\,\mathrm{AU}$. The very large value of the pulse period second derivative $\ddot{P} = -2.3 \times 10^{-27}\,\mathrm{s}^{-1}$ (Backer 1993), indicates the presence of a second, more distant orbital companion (Backer 1993; Backer, Foster, & Sallmen 1993; Thorsett, Arzoumanian, & Taylor 1993). The current timing data are consistent with a second companion mass in the range $m_2 \sim 10^{-3}$—$1\,M_\odot$, with a corresponding semi-major axis $a_2 \sim 10$—$100\,\mathrm{AU}$ for the outer orbit (Michel 1994a,b). Recently, a candidate optical counterpart consistent with being a main-sequence (hereafter MS) star of mass $m_2 \approx 0.5\,M_\odot$ has been identified by Bailyn et al. (1994). In this paper we assume that this optical identification is correct (and therefore we do not consider the possibility that the second companion might be a Jupiter-like planet; see Sigurdsson 1993), and we explore the consequences theoretically. Our results would not be changed significantly if the second companion were a MS star of somewhat lower mass. A MS star more massive than $0.5\,M_\odot$ is strictly ruled out by the optical observations.

The binary pulsar's orbital eccentricity, $e_1 = 0.025$, although very small, is several orders of magnitude larger than observed in most other low-mass binary millisecond pulsars (Thorsett et al. 1993). If this anomalous eccentricity has been induced by secular perturbations in the three-body system, then the second companion mass must be $\sim 0.1$—$1\,M_\odot$ (Rasio 1994a,b), consistent with the optical identification of Bailyn et al. (1994). Alternatively, the eccentricity might have been produced by the dynamical interaction that formed the system. We discuss this possibility briefly in §2.

Hierarchical triple systems are expected to be produced quite easily in globular clusters through dynamical interactions involving primordial binaries (Mikkola 1984; Hut 1992). There is growing evidence that the primordial binary fraction in globular clusters may be as large as 30%, with 10% generally considered "typical" (Hut et al. 1992b). When binaries dominate the core population (because of mass segregation), a few percent of them are expected to be in triples (McMillan, Hut, & Makino 1991). Here we consider in detail interactions between a wide primordial MS binary and a binary millisecond pulsar (hereafter BMP). In a significant fraction of these interactions, one of the two MS stars is ejected while the other remains bound in a wide orbit around the BMP. We note that interactions between a binary and a *single* star cannot produce a stable hierarchical triple in the absence of dissipation (Chazy 1929; but see Bailyn 1989).

Our motivation for invoking a pre-existing BMP is the following. If the second pulsar companion is indeed a MS star of mass $\approx 0.5\,M_\odot$, then the eccentricity of the outer orbit must be large ($e_2 > 0.5$) and its semi-major axis cannot be much smaller than $\sim 50$–$100\,\mathrm{AU}$, to be consistent with the pulsar timing data (Michel 1994a,b; Rasio 1994a,b). Such a wide orbit is easily disrupted in the dense core of M4 (at present PSR B1620$-$26 is most likely just outside the core but the density there is not significantly lower): its lifetime is then $\sim 10^7$–$10^8\,\mathrm{yr}$ (see §4). This is much shorter than the age of the BMP, which is most likely $\gtrsim 10^9\,\mathrm{yr}$ (see Thorsett et al. 1993). Thus the millisecond pulsar could not have formed inside the triple system.

In §2 and §3, we present the results of detailed numerical calculations of interactions between a BMP and a wide primordial MS binary. These calculations were done with the automated software package STARLAB (Hut 1994; Hut & McMillan 1994). The long-term, high accuracy triple integrations described in this paper are made possible by the use of a new time-symmetrized integration algorithm, whose remarkable stability properties are described in more detail by Hut, Makino, & McMillan (1994). In §4 we discuss the formation rates of triple systems like PSR B1620$-$26.

## 2. Eccentricity Perturbations

Since the eccentricity of the inner BMP in the presently observed system is very small, the BMP cannot have been perturbed very much during the exchange interaction[4]. This places a limit on how close either of the two MS stars can have approached the BMP. In our calculations of exchange interactions (§3), the BMP is treated as a *single mass* and given an effective radius $R_e$ such that close passages with periastron separation $r_p > R_e$ cannot produce an eccentricity larger than observed today. We first determine an appropriate value for $R_e$.

Fig. 1 shows the induced eccentricity $\Delta e_1$ as a function of pericenter distance. We assume that the BMP had a perfectly circular orbit before the encounter and we place the perturber on a slightly hyperbolic orbit with relative velocity at infinity $v_\infty \simeq 8\,\mathrm{km\,s^{-1}}$, comparable to the 3-D velocity dispersion in the cluster[5]. We see that the constraint $\Delta e_1 < 0.025$ (the value

---

[4]Tidal circularization is completely negligible given the large binary separation $a_1 \approx 0.8\,\mathrm{AU}$. Even if the pulsar's inner companion were a fully convective low-mass MS star, the eccentricity damping time would be $\sim 10^{18}\,\mathrm{yr}$ (using eq. [6.2] of Zahn 1977). In reality, the inner companion is most likely a white dwarf and the effect is even smaller.

[5]Because we are considering marginally soft primordial binaries, this is also the typical orbital velocity of

observed today) implies that no star of mass $0.5\,M_\odot$ can approach within $R_e \approx 3a_1 \simeq 2.4\,\mathrm{AU}$ of the BMP. This decreases to $R_e \approx 2.5a_1$ for a perturber of mass $0.2\,M_\odot$ and increases to $R_e \approx 3.5a_1$ for $0.8\,M_\odot$. In addition, for random orientations and phases, the distribution of $\Delta e_1$ values around the average is rather wide (and markedly non-Gaussian, with a long tail extending to small values of $\Delta e_1$; see Fig. 1, insert). Thus one cannot strictly define an effective radius $R_e$ that applies to all situations. However, given the enormous computational advantage of treating the BMP as a single mass, we nevertheless adopt this prescription as an approximation.

It is possible that a close encounter during the exchange interaction might have induced the eccentricity of 0.025 observed today. However, we find this possibility less attractive for the following reasons. First, because the dependence of $\Delta e_1$ on $r_p$ is very steep (close to the $\Delta e_1 \propto \exp(-r_p^{3/2})$ relation expected for small deviations from adiabatic invariance; see Phinney 1992, Heggie & Rasio 1994), one would have to tune the parameters quite carefully to obtain an eccentricity near the desired value. In addition, as shown by Rasio (1994a,b), a second companion of mass $m_2 \approx 0.5\,M_\odot$ can induce very naturally this measured eccentricity through secular orbital perturbations, so there is no real need for any other explanation.

## 3. Exchange Interactions

We now consider interactions between the BMP and a MS binary, and we determine the cross section for triple formation and the typical characteristics of the triples formed through this mechanism. We represent the BMP as a single mass with effective radius $R_e = 3a_1$. One of the two MS stars must be the present outer member of the triple, of mass $\approx 0.5\,M_\odot$. The other can have any mass but is most likely of mass $\lesssim 0.5\,M_\odot$ since the least massive component is most likely to be ejected during the interaction. For definiteness we have considered the three values 0.2, 0.5, 0.8 $M_\odot$.

Fig. 2 gives the total interaction cross section $\Sigma$ as a function of $a_{MS}$, the semi-major axis of the incoming MS binary. Each cross section was determined by simulating numerically $\sim 5 \times 10^4$ interactions. The statistical error on $\Sigma$ is always less than 5%. Energy equipartition and a Maxwellian distribution for the relative velocity $v_r$ have been assumed. The average $\langle \Sigma v_r \rangle$ shown in Fig. 2 is over this Maxwellian distribution. We adopt the value $\sigma = 5\,\mathrm{km\,s^{-1}}$

---

a MS star during the exchange interaction. We did not attempt to perform an average over relative velocities since the distribution of orbital velocities during the interaction is not known, and, in particular, need not be Maxwellian. The results are not sensitive to the adopted value for $v_\infty$, since the trajectories are never far from parabolic.

for the 1-D velocity dispersion of $0.8\,M_\odot$ objects. For interactions between objects of mass $m_a$ and $m_b$ and 1-D velocity dispersion $\sigma_a$ and $\sigma_b$, the relative (3-D) velocity dispersion is then given by $\sigma_r^2 = 3(\sigma_a^2 + \sigma_b^2)$ with $m_a\sigma_a^2 = m_b\sigma_b^2 = 0.8\,M_\odot\sigma^2$. The eccentricity of the primordial binaries is assumed to follow a thermal distribution, with probability density $p(e) = 2e$ (Heggie 1975). We see that in the region of interest we have $\langle \Sigma_{\rm tot} v_r \rangle \approx A R_e a_{MS} \sigma_r$, with the constant $A \approx 200$.

Fig. 2 also shows the branching ratios for different types of outcome. These are labelled "I" for ionization (i.e., unbinding the MS binary without triple formation), "E" for a "non-merging" exchange (triple formation with no star approaching within $R_e$ of the BMP), with either of the two MS stars ejected, and "M" for a "merger" (at least one star approaching within $R_e$ of the BMP). For two $0.5\,M_\odot$ MS stars, type M dominates when $a_{MS} \lesssim 8\,\mathrm{AU}$, and type I dominates for $a_{MS} \gtrsim 25\mathrm{AU}$. The formation of a triple system like PSR B1620−26, corresponding to an outcome of type E, is dominant in the interval $8\,\mathrm{AU} \lesssim a_{MS} \lesssim 25\,\mathrm{AU}$, with the branching ratio reaching a maximum of about 50% for $a_{MS} \approx 12\,\mathrm{AU}$.

In Figs. 3 and 4 we show the probability distributions of $a_2$ and $e_2$, the semi-major axis and eccentricity of the outer orbit in the triple system for several typical cases. Clearly, our mechanism is capable of forming triples with the desired characteristics: a circular inner orbit (that remains unperturbed after the interaction) and a very eccentric outer orbit with $a_2 \sim 10^2\,\mathrm{AU}$. We find that the average $\langle a_2 \rangle / a_{MS} \approx 3 - 5$, but with a wide distribution extending over more than an order of magnitude. The eccentricity distribution peaks at some value $e_2 \approx 0.6$ for tight incoming binaries with $a_{MS} \lesssim 5\,\mathrm{AU}$. For very wide incoming binaries with $a_{MS} > 20 AU$, the distribution has a sharp peak at $e_2 = 1$. For intermediate values of $a_{MS}$, the distribution of $e_2$ is close to thermal (and this is independent of the assumed distribution of eccentricities for the incoming MS binaries; see Fig. 4).

## 4. Timescales and Triple Formation Rate

Clearly, our scenario requires the presence in M4 of a significant population of wide primordial binaries with separations in the range $5 - 50\,\mathrm{AU}$. For comparison, the maximum separation for a hard binary is $a_{\rm soft} \equiv Gm/2\sigma^2 = 10\,\mathrm{AU}\,m_{0.5}\sigma_5^{-2}$, where $m_{0.5}$ is the average stellar mass in units of $0.5\,M_\odot$ and $\sigma_5$ is the 1-D velocity dispersion in units of $5\,\mathrm{km\,s^{-1}}$. The most recent measurements of the velocity dispersion in M4 (Peterson, Rees, & Cudworth 1994) give a projected velocity dispersion $\sigma_c \approx 3.5\,\mathrm{km\,s^{-1}}$ in the core, decreasing to $\sigma_h \approx 3\,\mathrm{km\,s^{-1}}$ at the half-mass radius. Thus primordial binaries with separations as large as $25\,\mathrm{AU}$ may in fact be (marginally) hard during most of their secular evolution outside the core. Even somewhat softer primordial binaries are not expected to be disrupted until after they have drifted into the cluster core. This is because the characteristic time for a binary to drift into the core is the two-body relaxation time, which is typically larger than the binary disruption time by a factor of order the Coulomb logarithm $\ln \Lambda \approx 10$ for a marginally soft binary. Using eq. (5.12) of Hut & Bahcall (1983) for the binary disruption rate, we find that the ratio of binary disruption time to relaxation time is given (to within $\approx 20\%$) by[6]

$$\frac{t_{\rm dis}}{t_{\rm relax}} \approx 10\, F(\mathcal{E}) \qquad \text{where } F(\mathcal{E}) \equiv \mathcal{E}^2 \left(1 + \frac{4}{15\mathcal{E}}\right)\left(1 + e^{-3\mathcal{E}/4}\right), \qquad (1)$$

where $\mathcal{E} \equiv Gm/(2a\sigma^2)$ measures the binary hardness. Note that this expression is independent of the local density. Eq. (1) becomes unity at $\mathcal{E} \approx 0.1$, or a critical separation $a_{\rm crit} \approx 10 a_{\rm soft} \approx 100\,{\rm AU}\, m_{0.5} \sigma_5^{-2}$. Thus primordial binaries with separations as large as $\sim 100\,{\rm AU}$ may be able to drift into the cluster core before they are disrupted.

The rate at which these wide primordial binaries drift into the cluster core today is $\sim N_b/t_d$, where $N_b$ is the number of wide primordial binaries still present in the outer region of the cluster and $t_d \sim 10^{10}$ yr (cf. Hut, McMillan, & Romani 1992a). We can write $N_b = f_b N_o$ where $f_b \sim 0.1$ is the binary fraction and $N_o \sim 10^5$. Soon after entering the core, the binary is disrupted in an interaction with another star or hard binary. A fraction $\beta_T \ll 1$ of these interactions is with a (detectable) BMP and leads to the formation of a system like PSR B1620−26. Crudely we can say that $\beta_T \lesssim N_{\rm BMP}/N_c$, where $N_{\rm BMP}$ is the number of detectable BMPs in the core (clearly $N_{\rm BMP} = 1$ in M4) and $N_c$ is the total number of objects in the core ($N_c \gtrsim 10^3$ for M4). We can write the average number $N_T$ of triples containing a detectable BMP at any given time as $N_T \sim (\beta_T N_b/t_d) t_T$, where $t_T$ is the average lifetime of such triples in the core[7]. Using again eq. (5.12) of Hut & Bahcall (1983) for a density $\rho = 10^4\, M_\odot\,{\rm pc}^{-3}$, average stellar mass $m = 1\, M_\odot$, and 1-D velocity dispersion $\sigma = 5\,{\rm km\,s}^{-1}$, we get $t_T \approx 2 \times 10^8\,{\rm yr}\, F(\mathcal{E})$, giving $t_T \sim 10^7$—$10^9$ yr for $\mathcal{E} \sim 0.1$—1, with $\mathcal{E} \approx 1.5\,(a_2/10\,{\rm AU})^{-1}$. Keeping all factors we obtain the following estimate:

$$N_T \sim 0.1\, N_{\rm BMP}\, \frac{f_b}{0.1} \left(\frac{N_c}{10^3}\right)^{-1} \left(\frac{N_o}{10^5}\right) \left(\frac{t_d}{10^{10}{\rm yr}}\right) \left(\frac{t_T}{10^8{\rm yr}}\right). \qquad (2)$$

---

[6]Eq. (1) takes into account only the disruption of the binary in a single strong interaction. The cumulative effect of distant encounters is not taken into account and may slightly reduce the value of $a_{crit}$. For very soft binaries ($\mathcal{E} \ll 1$), this reduction would be by a factor $\ln \Lambda \approx 10$, so that $a_{\rm crit} \approx a_{\rm soft}$ in that case (cf. Binney & Tremaine 1987, §8.4). However, for marginally soft binaries with $\mathcal{E} \lesssim 1$, the results of Hut (1983) indicate that the reduction is by a factor of 2 at most.

[7]Note that the typical recoil velocity of the triple is negligible here and so the triple is not expected to spend any time far outside the core after it forms. The position of PSR B1620−26 is projected inside, but most likely just outside, the core of M4.

Given the large uncertainties, this estimate seems perfectly compatible with the detection of just one triple system in M4. In addition, we note that there are now $\sim 10$ BMPs known in all globular clusters, but only one (PSR B1620−26 ) appears to be in a triple.

We acknowledge gratefully the hospitality of the ITP at UC Santa Barabara, where this work was begun. F. A. R. is supported by a Hubble Fellowship, funded by NASA through Grant HF-1037.01-92A from the Space Telescope Science Institute, which is operated by AURA, Inc., under contract NAS5-26555. Partial support for this work was also provided by NASA Grant NAGW-2559 and NSF Grant AST-9308005.

Fig. 1.— Average eccentricity induced in the orbit of the binary millisecond pulsar (BMP) after an interaction with a passing star of mass $0.5\,M_\odot$ with a distance of closest approach $r_p$ to the BMP's center of mass (solid square dots). The average is over all relative inclinations, orientations, and phases (assumed random). The relative velocity at infinity has a fixed value $v_\infty \simeq 8\,\mathrm{km\,s^{-1}}$. The error bars show the spread between the 10th and 90th percentiles in the distributions. A typical distribution, corresponding to $r_p/a_1 = 3$, is shown in the insert. For comparison, we also show the average eccentricity induced by a passing star of mass $0.2\,M_\odot$ (open square dots) and $0.8\,M_\odot$ (open triangles). The present value of the BMP's eccentricity, $e_1 = 0.025$, is shown by the dashed lines. Interactions with $r_p < R_e \approx 3a_1$ would produce an eccentricity larger than observed today.

Fig. 2.— (a) Total cross section for the interaction between the BMP and a MS binary with semi-major axis $a_{MS}$ containing two $0.5\,M_\odot$ stars. The "effective radius" of the BMP is $R_e = 3a_1$, where $a_1$ is the BMP's semi-major axis. (b) Branching ratios for various types of outcome: I= ionization; E= exchange; M= merger (see text for details). The thinner lines show how the results vary when some of the parameters are changed: increasing $R_e$ to $4a_1$ (dotted lines), increasing one of the MS masses to $0.8\,M_\odot$ (short-dashed lines), or decreasing it to $0.2\,M_\odot$ (long-dashed lines). The formation of a triple system like PSR B1620−26 , corresponding here to an outcome of type E, dominates the cross section for $a_{MS} \approx 10\,\mathrm{AU}$.

Fig. 3.— Distribution of the semi-major axis $a_2$ of the triple's outer orbit (above) and distribution in the $(a_2, e_2)$ plane (below). The size of each square dot is proportional to the differential cross-section for producing a triple system with specific orbital parameters. Here the incoming MS binary contained two $0.5\,M_\odot$ stars with semi-major axis $a_{MS} \simeq 13\,\mathrm{AU}$.

Fig. 4.— Distribution of the eccentricity $e_2$ of the triple's outer orbit. The incoming MS binary contained two $0.5\,M_\odot$ stars. Three values of the semi-major axis are shown, $a_{MS} \simeq 3\,\mathrm{AU}$ (a), $13\,\mathrm{AU}$ (b), and $50\,\mathrm{AU}$ (c). For reference, the thin straight line shows a thermal eccentricity distribution $p(e) = 2e$. The eccentricity of the incoming MS binaries was assumed to follow a thermal distribution in all cases, except (b) where we also show the result corresponding to circular incoming binaries (dotted line).